\documentclass{amsart}
\usepackage{hyperref}
\newcommand{\arxiv}[1]{\href{http://arxiv.org/abs/#1}{arXiv:#1}}
\newcommand*{\mailto}[1]{\href{mailto:#1}{\nolinkurl{#1}}}
\makeatletter
\def\bibitem{\@ifnextchar [{\@ltestbibitem}{\@testbibitem}}
\def\@testbibitem#1{\@ifnextchar ({\@mrbibitem{#1}}{\@bibitem{#1}}}
\def\@ltestbibitem[#1]#2{\@ifnextchar ({\@mrbibitem{#2}}{\@bibitem{#2}}}
\def\@mrbibitem#1(#2){\@ifundefined{href}{\item}{\stepcounter{\@listctr}\item[{[\href{http://www.ams.org/mathscinet-getitem?mr=#2&return=pdf}{\the\value{\@listctr}}]}]\Hy@raisedlink{\hyper@anchorstart {cite.#1}\relax \hyper@anchorend}}\if@filesw\immediate\write\@auxout{\string\bibcite{#1}{\the\value{\@listctr}}}\fi\ignorespaces}
\def\@lmrbibitem#1#2(#3){\item[\@ifundefined{href}{\@biblabel{#1}}{\@ifnotempty{#1}{[\href{http://www.ams.org/mathscinet-getitem?mr=#3&return=pdf}{#1}]}}\hfill]\if@filesw{\let\protect\noexpand\immediate\write\@auxout{\string\bibcite{#2}{#1}}}\fi\ignorespaces}
\makeatother

\newtheorem{theorem}{Theorem}[section]
\newtheorem{lemma}[theorem]{Lemma}

\newtheorem{hypo}[theorem]{Hypothesis {\bf H.}\hspace*{-0.6ex}}

\newcommand{\R}{{\mathbb R}}
\newcommand{\N}{{\mathbb N}}
\newcommand{\Z}{{\mathbb Z}}

\newcommand{\nn}{\nonumber}
\newcommand{\be}{\begin{equation}}
\newcommand{\ee}{\end{equation}}

\newcommand{\ti}{\tilde}

\newcommand{\spr}[2]{\langle #1 , #2 \rangle}
\newcommand{\id}{{\rm 1\hspace{-0.6ex}l}}
\newcommand{\E}{\mathrm{e}}

\newcommand{\lz}{\ell^2(\Z)}
\newcommand{\tl}{\mathrm{TL}}
\newcommand{\km}{\mathrm{KM}}

\newcommand{\eps}{\varepsilon}

\numberwithin{equation}{section}

\begin{document}

\title{On the Spatial Asymptotics of Solutions of the Toda Lattice}

\author[G. Teschl]{Gerald Teschl}
\address{Faculty of Mathematics\\ University of Vienna\\
Nordbergstrasse 15\\ 1090 Wien\\ Austria\\ and International Erwin Schr\"odinger
Institute for Mathematical Physics\\ Boltzmanngasse 9\\ 1090 Wien\\ Austria}
\email{\mailto{Gerald.Teschl@univie.ac.at}}
\urladdr{\url{http://www.mat.univie.ac.at/~gerald/}}

\thanks{Work supported by the Austrian Science Fund (FWF) under Grant No.\ Y330.}
\thanks{Discrete Contin. Dyn. Syst. {\bf 27:3}, 1233--1239 (2010)}

\keywords{Toda lattice, spatial asymptotics, Toda hierarchy, Kac--van Moerbeke hierarchy}
\subjclass[2000]{Primary 37K40, 37K15; Secondary 35Q53, 37K10}

\begin{abstract}
We investigate the spatial asymptotics of decaying solutions of the Toda lattice and show that
the asymptotic behavior is preserved by the time evolution. In particular, we show that the leading
asymptotic term is time independent. Moreover, we establish infinite propagation speed for the
Toda lattice. All results are extended to the entire Toda as well as the Kac--van Moerbeke hierarchy.
\end{abstract}

\maketitle

\section{Introduction}

Since the seminal work of Gardner et al.\ \cite{ggkm} in 1967 it is known that completely integrable wave
equations can be solved by virtue of the inverse scattering transform. In particular, this implies
that short-range perturbations of the free solution remain short-range during the time evolution. This
raises the question to what extend spatial asymptotical properties are preserved during
time evolution. In \cite{bo}, \cite{bs} (see also \cite{kpst}) Bondareva and Shubin considered
the initial value problem for the Korteweg--de Vries (KdV) equation in the class of initial conditions which
have a prescribed asymptotic expansion in terms of powers of the spatial variable. As part of their analysis they
obtained that the leading term of this asymptotic expansion is time independent. Inspired by this
intriguing fact, the aim of the present paper is to prove a general result for the Toda equation which
contains the analog of this result plus the known results  for short-range perturbation alluded to before as
a special case.

More specifically, recall the Toda lattice \cite{ta} (in Flaschka's variables \cite{fl1})
\begin{align} \nn
\frac{d}{dt} a(n,t) &= a(n,t) \Big(b(n+1,t)-b(n,t)\Big), \\ \label{todeqfl}
\frac{d}{dt} b(n,t) &= 2 \Big(a(n,t)^2-a(n-1,t)^2\Big), \qquad n\in \Z.
\end{align}
It is a well studied physical model and the prototypical discrete integrable wave equation.
We refer to the monographs \cite{fad}, \cite{ghmt}, \cite{tjac}, \cite{ta} or the review articles \cite{krt}, \cite{taet}
for further information.

Then our main result, Theorem~\ref{thmmain} below, implies for example that
\be\label{exasym}
a(n,t) = \frac{1}{2} + \frac{\alpha}{n^\delta} + O(\frac{1}{n^{\delta+\eps}}), \quad
b(n,t) = \frac{\beta}{n^\delta} + O(\frac{1}{n^{\delta+\eps}}), \quad n\to\infty,
\ee
for all $t\in\R$ provided this holds for the initial condition $t=0$. Here $\alpha,\beta\in\R$
and $\delta\ge 0$, $0<\eps\le 1$. 

A few remarks are in order: First of all, it is important to point out that the error
terms will in general grow with $t$ (see the discussion after Theorem~\ref{thmmain} for
a rough time dependent bound on the error). An analogous result holds for $n\to-\infty$. Moreover,
there is nothing special about the powers $n^{-\delta}$, which can be replaced by
any bounded sequence which, roughly speaking, does decay at most exponentially
and whose difference is asymptotically of lower order. Finally, similar results hold for the
Ablowitz--Ladik equation. However, since the Ablowitz--Ladik system does not have
the same difference structure some modifications are neccessary and will be given
in Michor \cite{m}.

\section{The Cauchy problem for the Toda lattice}

To set the stage let us recall some basic facts for the Toda lattice. We will only consider
bounded solutions and hence require

\begin{hypo} \label{habt}
Suppose $a(t)$, $b(t)$ satisfy
\[
a(t) \in \ell^{\infty}(\Z, \R), \qquad b(t) \in \ell^{\infty}(\Z, \R), \qquad
a(n,t) \neq 0, \qquad (n,t) \in \Z \times \R,
\]
and let $t \mapsto (a(t), b(t))$ be differentiable in 
$\ell^{\infty}(\Z) \oplus \ell^{\infty}(\Z)$.
\end{hypo}

First of all, to see complete integrability it suffices to find a so-called Lax pair \cite{lax}, that
is, two operators $H(t)$, $P(t)$ in $\lz$ such that the Lax equation
\begin{equation} \label{laxeq}
\frac{d}{dt} H(t) = P(t) H(t) - H(t) P(t)
\end{equation}
is equivalent to \eqref{todeqfl}. Here $\lz$ denotes the Hilbert space of square summable
(complex-valued) sequences over $\Z$. One can easily convince oneself that the right
choice is
\begin{align} \nn
H(t) &= a(t) S^+ + a^-(t) S^- + b(t),\\
P(t) &= a(t) S^+ - a^-(t) S^-,
\end{align}
where $(S^\pm f)(n) = f^\pm(n)= f(n\pm1)$ are the usual shift operators.

Now the Lax equation \eqref{laxeq} implies that the operators $H(t)$ for
different $t\in\R$ are unitarily equivalent (cf.\ \cite[Thm.~12.4]{tjac}):

\begin{theorem}\label{thmunitary}
Let $P(t)$ be a family of bounded skew-adjoint operators, such that $t\mapsto
P(t)$ is differentiable. Then there exists a family of unitary propagators $U(t,s)$
for $P(t)$, that is,
\begin{equation}
\frac{d}{dt} U(t,s) = P(t) U(t,s), \qquad U(s,s)=\id.
\end{equation}
Moreover, the Lax equation \eqref{laxeq} implies
\begin{equation}
H(t)= U(t,s) H(s) U(t,s)^{-1}.
\end{equation}
\end{theorem} 

As pointed out in \cite{ttkm}, this result immediately implies global existence of bounded solutions of the
Toda lattice as follows:
Considering the Banach space of all bounded real-valued coefficients $(a(n),b(n))$ (with the sup norm),
local existence is a consequence of standard results for differential equations in Banach spaces. Moreover,
Theorem~\ref{thmunitary} implies that the norm $\|H(t)\|$ is constant, which in turn
provides a uniform bound on the coefficients of $H(t)$,
\begin{equation}\label{abinfh}
\|a(t)\|_\infty + \|b(t)\|_\infty \le 2\|H(t)\| = 2\|H(0)\|.
\end{equation}
Hence solutions of the Toda lattice cannot blow up and are global in time (see \cite[Sect.~12.2]{tjac}
for details):

\begin{theorem} \label{thmexistandunique}
Suppose $(a_0,b_0) \in M = \ell^\infty(\Z,\R) \oplus \ell^\infty(\Z,\R)$. Then there exists a unique
integral curve $t \mapsto (a(t),b(t))$ in $C^\infty(\R,M)$ of the Toda lattice \eqref{todeqfl} such
that $(a(0),b(0)) = (a_0,b_0)$.
\end{theorem}

However, more can be shown. In fact, when considering the inverse scattering transform for the Toda lattice it
is desirable to establish existence of solutions within the Marchenko class, that is, solutions satisfying
\begin{equation}
\sum_{n\in\Z} (1+|n|) \Big( |a(n,t)-\frac{1}{2}| + |b(n,t)| \Big) <\infty
\end{equation}
for all $t\in\R$. That this is indeed true was first established in \cite{tist} and rediscovered in \cite{kha} using
a different method.
Furthermore, the weight $1+|n|$ can be replaced by an
(almost) arbitrary weight function $w(n)$.

\begin{lemma}\label{lemscat}
Suppose $a(n,t)$, $b(n,t)$ is some bounded solution of the Toda lattice \eqref{todeqfl} satisfying
\eqref{decay} for one $t_0 \in \R$. Then
\be \label{decay}
\sum_{n \in \Z} w(n) \Big(|a(n,t) - \frac{1}{2}| + |b(n,t)| \Big) 
< \infty,
\ee
holds for all $t \in \R$, where $w(n)\ge 1$ is some weight with $\sup_n( |\frac{w(n+1)}{w(n)}| + |\frac{w(n)}{w(n+1)}| )<\infty$.
\end{lemma}

Moreover, as was demonstrated in \cite{emtist} (see also \cite{emtist2}), one can even replace $|a(n,t) - \frac{1}{2}| + |b(n,t)|$
by $|a(n,t) - \bar a(n,t)| + |b(n,t) - \bar b(n,t)|$, where $\bar a(n,t)$, $\bar b(n,t)$ is some other bounded
solution of the Toda lattice. See also \cite{kha2}, where similar results are shown.

This result shows that the asymptotic behavior as $n\to\pm\infty$ is preserved to leading order by the Toda lattice.
The purpose of this paper is to show that even the leading term is preserved (i.e., time independent) by
the time evolution.

Set
\be
\|(a,b)\|_{w,p} = \begin{cases}
\left(\sum\limits_{n \in \Z} w(n) \Big(|a(n)|^p + |b(n)|^p \Big)\right)^{1/p}, & 1\le p <\infty\\
\sup\limits_{n \in \Z} w(n) \Big(|a(n)| + |b(n)| \Big), & p=\infty.
\end{cases}
\ee
Then one has the following result:

\begin{theorem}\label{thmmain}
Let $w(n)\ge 1$ be some weight with $\sup_n( |\frac{w(n+1)}{w(n)}| + |\frac{w(n)}{w(n+1)}| )<\infty$ and
fix some $1 \le p \le \infty$.
Suppose $a_0, b_0$ and $\ti{a}_0$, $\ti{b}_0$ are bounded sequences such that
\be
\|(a_0^+-a_0, b_0^+-b_0)\|_{w,p} < \infty \quad\text{and}\quad \|(\ti{a}_0,\ti{b}_0)\|_{w,p} < \infty.
\ee
Suppose $a(t), b(t)$ is the unique solution of the Toda lattice \eqref{todeqfl} corresponding to the
initial conditions
\be
a(0) = a_0 + \ti{a}_0\ne 0, \quad b(0) = b_0 + \ti{b}_0.
\ee
Then this solution is of the form
\be
a(t) = a_0 + \ti{a}(t), \quad b(t) = b_0 + \ti{b}(t), \quad\text{where}\quad \|(\ti{a}(t),\ti{b}(t))\|_{w,p} < \infty
\ee
for all $t\in\R$.
\end{theorem}

\begin{proof}
The Toda equation \eqref{todeqfl} implies the differential equation
\begin{align} \nn
\frac{d}{dt} \ti{a}(n,t) =& a(n,t) \Big(\ti{b}(n+1,t)-\ti{b}(n,t) + b_0(n+1)-b_0(n)\Big), \\ \nn
\frac{d}{dt} \ti{b}(n,t) =& 2 \Big(\big(a(n,t)+a_0(n)\big) \ti{a}(n,t) - \big(a(n-1,t)+a_0(n-1)\big) \ti{a}(n-1,t)\\ \label{todeqflti}
& +(a_0(n)+a_0(n-1))(a_0(n)-a_0(n-1)) \Big), \qquad n\in \Z
\end{align}
for $(\ti{a},\ti{b})$. Since our requirement for $w(n)$ implies that the shift operators are continuous with respect to
the norm $\|.\|_{w,p}$ and the same is true for the multiplication operator with a bounded sequence, this is an
inhomogeneous linear differential equation in our Banach space which has a unique global solution in this
Banach space (e.g., \cite[Sect.~1.4]{dei}). Moreover, since $w(n)\ge 1$ this solution is bounded and the
corresponding coefficients $(a,b)$ coincide with the solution of the Toda equation from
Theorem~\ref{thmexistandunique}.
\end{proof}

Note that using Gronwall's inequality one can easily obtain an explicit bound
\be
\|(\ti{a}(t),\ti{b}(t))\|_{w,p} \le \|(\ti{a}_0(t),\ti{b}_0(t))\|_{w,p} \E^{C t} +
\|(a_0^+-a_0,b_0^+-b_0)\|_{w,p} \frac{1}{C} ( \E^{C t} -1),
\ee
where $C= 4(\|H\| + \|a_0\|_\infty)$ (since $\|a(t)\|_\infty\le \|H\|$ by \eqref{abinfh}).

To see the claim \eqref{exasym} from the introduction, let
\be
a_0(n) = \frac{1}{2} + \frac{\alpha}{n^\delta}, \quad b_0(n) = \frac{\beta}{n^\delta}, \quad \alpha,\beta\in\R, \delta>0,
\ee
for $n>0$ and $a_0(n)=b_0(n)=0$ for $n\le 0$. Now choose $p=\infty$ with
\be
w(n) = \begin{cases} (1+|n|)^{\delta+\eps}, & n>0,\\ 1, & n \le 0. \end{cases}
\ee
and apply the previous theorem. To see Lemma~\ref{lemscat}, just choose $a_0(n)=\frac{1}{2}$, $b_0(n)=0$ and
$p=1$.

Finally, let us remark that the requirement that $w(n)$ does not grow faster than exponentially is important.
If it were not present, our result would imply that a compact perturbation of the free solution $a(n,t)=\frac{1}{2}$,
$b(n,t)=0$ remains compact for all time if and only if it is equal to the free solution. This is well-known for the
KdV equation \cite{zh}, but we are not aware of a reference for the Toda equation.

\begin{theorem}
Let $a(n,t)$, $b(n,t)$ be a bounded solution of the Toda lattice \eqref{todeqfl}. If the sequences $a(n,t)-\frac{1}{2}$,
$b(n,t)$ are zero for all except for a finite number of $n\in\Z$ for two different times $t_0\ne t_1$, then they vanish
identically.
\end{theorem}

\begin{proof}
Without loss we can choose $t_0=0$ and suppose that the sequences $a(n,0)-\frac{1}{2}$, $b(n,0)$ are zero for
all except for a finite number of $n$. Then the associated reflection coefficients $R_\pm(k,0)$ (see
\cite{tjac} Chapter~10) are rational functions with respect to $k$ and by the inverse scattering transform
(\cite{tjac} Theorem~13.8) we have $R_\pm(k,t) = R_\pm(k,0) \exp(\pm (k-k^{-1}) t)$, which is not rational for any
$t\ne 0$ unless $R_\pm(k,t) \equiv 0$. Hence it must be a pure $N$ soliton solution, which has compact support
if and only if it is trivial, $N=0$.
\end{proof}

For related unique continuation results for the Toda equation see Kr\"uger and Teschl \cite{krt2}.

\section{Extension to the Toda and Kac--van Moerbeke hierarchy}
\label{secth}

In this section we show that our main result extends to the entire Toda hierarchy (which will
cover the Kac--van Moerbeke hierarchy as well).
To this end, we introduce the Toda hierarchy using the standard Lax formalism
following \cite{bght} (see also \cite{ghmt}, \cite{tjac}).

Choose constants $c_0=1$, $c_j$, $1\le j \le r$, $c_{r+1}=0$, and set
\begin{align} 
\begin{split}
g_j(n,t) &= \sum_{\ell=0}^j c_{j-\ell} \spr{\delta_n}{H(t)^\ell \delta_n},\\ \label{todaghsp}
h_j(n,t) &= 2 a(n,t) \sum_{\ell=0}^j c_{j-\ell}  \spr{\delta_{n+1}}{H(t)^\ell
\delta_n} + c_{j+1}.
\end{split}
\end{align}
The sequences $g_j$, $h_j$ satisfy the recursion relations
\begin{align}
\nn g_0 = 1, \: h_0 &= c_1,\\ \nn
2g_{j+1} -h_j -h_j^- -2b g_j &= 0,\quad 0 \le j\le r,\\ \label{rectodah}
h_{j+1} -h_{j+1}^- - 2(a^2 g_j^+ -(a^-)^2 g_j^-) - b
(h_j -h_j^-) &= 0, \quad 0 \le j < r.
\end{align}
Introducing
\begin{equation}  \label{btgptdef}
P_{2r+2}(t) = -H(t)^{r+1} + \sum_{j=0}^r ( 2a(t) g_j(t) S^+ -h_j(t)) H(t)^{r-j} +
g_{r+1}(t),
\end{equation}
a straightforward computation shows that the Lax equation
\begin{equation} \label{laxp}
\frac{d}{dt} H(t) -[P_{2r+2}(t), H(t)]=0, \qquad t\in\R,
\end{equation}
is equivalent to
\begin{equation}\label{tlrabo}
\tl_r (a(t), b(t)) = \begin{pmatrix} \dot{a}(t) -a(t) \Big(g_{r+1}^+(t) -
g_{r+1}(t) \Big)\\ 
\dot{b}(t) - \Big(h_{r+1}(t) -h_{r+1}^-(t) \Big) \end{pmatrix} =0,
\end{equation}
where the dot denotes a derivative with respect to $t$.
Varying $r\in \N_0$ yields the Toda hierarchy $\tl_r(a,b) =0$.

All results mentioned in the previous section, Theorem~\ref{thmunitary}, Theorem~\ref{thmexistandunique},
and Lemma~\ref{lemscat} remain valid for the entire Toda hierarchy (see \cite{tjac}) and so does our main result:

\begin{theorem}
Let $w(n)\ge 1$ be some weight with $\sup_n( |\frac{w(n+1)}{w(n)}| + |\frac{w(n)}{w(n+1)}| )<\infty$ and
fix some $1 \le p \le \infty$.
Suppose $a_0, b_0$ and $\ti{a}_0$, $\ti{b}_0$ are bounded sequences such that
\be
\|(a_0^+-a_0, b_0^+-b_0)\|_{w,p} < \infty \quad\text{and}\quad \|(\ti{a}_0,\ti{b}_0)\|_{w,p} < \infty.
\ee
Suppose $a(t), b(t)$ is the unique solution of some equation of the Toda hierarchy, $\tl_r(a,b) =0$, corresponding to the
initial conditions
\be
a(0) = a_0 + \ti{a}_0>0, \quad b(0) = b_0 + \ti{b}_0.
\ee
Then this solution is of the form
\be
a(t) = a_0 + \ti{a}(t), \quad b(t) = b_0 + \ti{b}(t), \quad\text{where}\quad \|(\ti{a}(t),\ti{b}(t))\|_{w,p} < \infty
\ee
for all $t\in\R$.
\end{theorem}

\begin{proof}
The proof is almost identical to the one of Theorem~\ref{thmmain}. From $\tl_r(a,b) =0$ one obtains an
inhomogeneous differential equation for $(\ti{a},\ti{b})$. The homogenous part is a finite sum over shifts
of $(\ti{a},\ti{b})$ and the inhomogeneous part is $\big(a_0 (g_{0,r+1}^+ -g_{0,r+1}(t)),
h_{0,r+1} -h_{0,r+1}^-\big)$, where $g_{0,r+1}$, $h_{0,r+1}$ are formed from $(a_0,b_0)$.
Finally, it is straightforward to show that the $\|.\|_{w,p}$ norm of the inhomogeneous part is finite
by induction using the recursive definition of $g_{r+1}(t)$ and $h_{r+1}(t)$.
\end{proof}

Similarly we also obtain

\begin{theorem}
Let $a(n,t)$, $b(n,t)$ be a bounded solution of the of some equation of the Toda hierarchy, $\tl_r(a,b) =0$.
If the sequences $a(n,t)-\frac{1}{2}$, $b(n,t)$ are zero for all except for a finite number of $n\in\Z$ for two
different times $t_0\ne t_1$, then they vanish identically.
\end{theorem}

Finally since the Kac--van Moerbeke hierarchy can be obtained by setting $b=0$ in the odd equations of the
Toda hierarchy, $\km_r(a) = \tl_{2r+1}(a,0)$ (see \cite{mt}), this last result also coveres the Kac--van Moerbeke hierarchy.

\section*{Acknowledgments}
I want to thank Ira Egorova, Fritz Gesztesy, Thomas Kappeler, and Helge Kr\"uger for discussions on this topic.


\begin{thebibliography}{XX}
\bibitem{bo} (MR0717670)
I. N. Bondareva, {\em The Korteweg--de Vries equation in classes of increasing functions with prescribed
asymptotic behavior as $|x|\to\infty$}, Mat. USSR Sb. {\bf 50:1}, 125--135 (1985).
\bibitem{bs} (MR0685832)
I. Bondareva and M. Shubin, {\em Increasing asymptotic solutions of the Korteweg--de Vries equation and its
higher analogues}, Sov. Math. Dokl. {\bf 26:3}, 716--719 (1982).
\bibitem{bght} (MR1432141)
W. Bulla, F. Gesztesy, H. Holden, and G. Teschl, {\em
Algebro-Geometric Quasi-Periodic Finite-Gap Solutions of the Toda and Kac-van
Moerbeke Hierarchies}, Mem. Amer. Math. Soc. {\bf 135-641}, (1998).
\bibitem{dei} (MR0463601)
K. Deimling, {\em Ordinary Differential Equations on Banach Spaces}, 
Lecture Notes in Mathematics {\bf 596}, Springer, Berlin, 1977.
\bibitem{emtist} (MR2286092)
I. Egorova, J. Michor, and G. Teschl,
{\em Inverse scattering transform for the Toda hierarchy with quasi-periodic background},
Proc. Amer. Math. Soc. {\bf 135}, 1817--1827 (2007).
\bibitem{emtist2} I. Egorova, J. Michor, and G. Teschl,
{\em Inverse scattering transform for the Toda hierarchy with steplike finite-gap backgrounds},
J. Math. Phys. {\bf 50}, 103521 (2009). 
\bibitem{fad} (MR0905674)
L. Faddeev and L. Takhtajan, {\em Hamiltonian Methods in the
Theory of Solitons}, Springer, Berlin, 1987.
\bibitem{fl1} (MR0408647)
H. Flaschka, {\em The Toda lattice. I. Existence of integrals}, Phys. Rev. B {\bf 9},
1924--1925 (1974).
\bibitem{ggkm}
C. S. Gardner, J. M. Green, M. D. Kruskal, and R. M. Miura,
{\em A method for solving the Korteweg-de Vries equation}, Phys. Rev.
Letters {\bf 19}, 1095--1097 (1967).
\bibitem{ghmt} (MR2446594)
F. Gesztesy, H. Holden, J. Michor, and G. Teschl, {\em Soliton Equations
and Their Algebro-Geometric Solutions. Volume II: $(1+1)$-Dimensional Discrete Models},
Cambridge Studies in Advanced Mathematics {\bf 114}, Cambridge University Press, Cambridge, 2008.
\bibitem{kha}
A. Kh. Khanmamedov, {\em On the rapidly decreasing solution of the Cauchy problem for the Toda chain},
Theoret. and Math. Phys. {\bf 142:1}, 1--7 (2005).
\bibitem{kha2} (MR2409495)
A. Kh. Khanmamedov, {\em The solution of CauchyÕs problem for the Toda lattice with limit periodic
initial data}, Sb. Math. {\bf 199:3}, 449--458 (2008).
\bibitem{kpst} (MR2393267)
T. Kappeler, P. Perry, M. Shubin and P. Topalov, {\em Solutions of mKdV in classes of functions
unbounded at infinity}, J. Geom. Anal. {\bf 18}, 443--477 (2008).
\bibitem{krt} (MR2493113)
H. Kr\"uger and G. Teschl, {\em Long-time asymptotics for the Toda lattice for decaying initial data revisited},
Rev. Math. Phys. {\bf 21:1}, 61--109 (2009).
\bibitem{krt2}
H. Kr\"uger and G. Teschl, {\em Unique continuation for discrete nonlinear wave equations},
\arxiv{0904.0011}.
\bibitem{lax} (MR0235310)
P. D. Lax {\em Integrals of nonlinear equations of evolution and
solitary waves}, Comm. Pure and Appl. Math. {\bf 21}, 467--490 (1968).
\bibitem{m}
J. Michor, {\em On the spatial asymptotics of solutions of the Ablowitz--Ladik hierarchy},
\arxiv{0909.3372}.
\bibitem{mt}
J. Michor and G. Teschl, {\em On the equivalence of different Lax pairs for the Kac-van Moerbeke hierarchy},
in Modern Analysis and Applications, V. Adamyan (ed.) et al., 445--453, Oper. Theory Adv. Appl. {\bf 191},
Birkh\"auser, Basel, 2009.
\bibitem{ttkm} (MR1703351)
G. Teschl, {\em On the Toda and Kac--van Moerbeke hierarchies}, Math. Z.
{\bf 231}, 325--344 (1999).
\bibitem{tist} (MR1694723)
G. Teschl, {\em Inverse scattering transform for the Toda hierarchy},
Math. Nach. {\bf 202}, 163--171 (1999).
\bibitem{tjac} (MR1711536)
G. Teschl, {\em Jacobi Operators and Completely Integrable Nonlinear Lattices},
Math. Surv. and Mon. {\bf 72}, Amer. Math. Soc., Rhode Island, 2000.
\bibitem{taet} (MR1879178)
G. Teschl, {\em Almost everything you always wanted to know about the
Toda equation}, Jahresber. Deutsch. Math.-Verein. {\bf 103}, no. 4, 149--162 (2001).
\bibitem{ta} (MR0971987)
M. Toda, {\em Theory of Nonlinear Lattices}, $2^{\text{nd}}$ enl.
edition, Springer, Berlin, 1989.
\bibitem{zh} (MR1145162)
B. Zhang, {\em Unique continuation for the Korteweg--de Vries equation}, SIAM J. Math. Anal. {\bf 23},
55--71 (1992).
\end{thebibliography}
\end{document}